# AI-assisted hyper-dimensional broadband quantum memory with efficiency above 90% in warm atoms

**Zeliang Wu[1], Jinxian Guo[2,7] ✉, Zhifei Yu[3], Wenfeng Huang[1], Chun-Hua Yuan[1,7] ✉, Weiping Zhang[2,4,5,6] & L. Q. Chen[1,4,7] ✉**

High-dimensional broadband quantum memory significantly expands quantum information processing capabilities, but the memory efficiency becomes insufficient when extended to high dimensions. We demonstrate an efficient quantum memory for hyper-dimensional photons encoded with orbital angular momentum (OAM) and spin angular momentum (SAM). OAM information is encoded from −5 to +5, combined with SAM encoding, enabling up to 22 dimensions. To ensure high memory efficiency, an artificial intelligence algorithm, a modified Differential Evolution (DE) algorithm using Chebyshev sampling, is developed to obtain a perfect signal-control waveform matching. Memory efficiency is experimentally achieved at 92% for single-mode Gaussian signal, 91% for information dimension of 6 and 80% for dimensional number to 22. The fidelity is achieved up to 99% for single-mode Gaussian signal, 95.5% for OAM information, 97.4% for SAM information, and 92% for whole hyper-dimensional signal, which is far beyond no-cloning limitation. Our results demonstrate superior performance and potential applications in high-dimensional quantum information processing. This achievement provides a crucial foundation for future quantum communication and quantum computing.

As the demand for distributed quantum computing[1–4] and long-distance quantum communication[5–7] grows, the requirement for high data throughput has become increasingly urgent in quantum networks[8,9]. High-dimensional broadband quantum memory is the performance metric for the scalability and effectiveness of quantum networks, ensuring that they can support the high throughput demanded by the evolving quantum landscape. A 50% efficiency is the baseline for functionality, surpassing the 90% threshold is crucial for practical application in quantum networks.

So far, many high-dimensional and broadband quantum states have been generated in high-dimensional spaces across various domains, such as time[10], spatial modes[11], and spectra[12–15] which are used for transmitting information. However, the efficiency of quantum memories remains a critical challenge, especially when scaling up to high dimensions and high bandwidths. Quantum states encoded with the OAM is a typical high-dimensional quantum field. Memories of OAM light have been demonstrated with the bandwidth below the MHz level[16,17], however, they exhibit a significant drop in efficiency with higher topological charges. The efficiency for an OAM state with a topological charge of $l = 1$ reaches 70%[18], but this efficiency decreases to 60% for $l = 12$, highlighting the challenge of maintaining efficiency with higher dimensions. Broadband quantum memory has been proven to have a bandwidth of up to 77 MHz and an efficiency of 82%, but it is only suitable for single-mode Gaussian signals[19]. Quantum memory integrating high efficiency, high dimensional, and broadband is still a challenge in the field of quantum information science.

In this article, we experimentally demonstrate efficient hyper-dimensional broadband quantum memory in an $^{87}$Rb atomic vapor cell using a far off-resonant Raman memory. To ensure high memory efficiency and high fidelity, we develop an artificial intelligence (AI) technology: a modified DE algorithm based on Chebyshev sampling, which enhances the manipulation accuracy of control waveforms. With the assistance of the DE algorithm, memory efficiency is achieved up to 92% and fidelity is 99% when the signal is a single-photon-level coherent optical pulse with the single-mode Gaussian (SMG) mode and bandwidth up to 50 MHz. Then, OAM and SAM are simultaneously encoded onto the broadband signal to obtain hyper-dimensional information. AI-assisted signal-control waveform matching technology results in memory efficiency of 91% for dimensionality up to 6 and gradually decrease to 80% when dimensionality to 22. The fidelity reaches 95.5% and 97.4% for the OAM and SAM information, and surpassing the no-cloning limitation when the memory times are 0.42 $\mu$s and 50 $\mu$s, respectively. These results meet the practical requirements of quantum memory, laying a solid technical foundation for large-scale and high-speed quantum networks.

[1]State Key Laboratory of Precision Spectroscopy, Quantum Institute for Light and Atoms, School of Physics and Electronic Science, East China Normal University, Shanghai, China. [2]School of Physics and Astronomy, and Tsung-Dao Lee Institute, Shanghai Jiao Tong University, Shanghai, China. [3]School of Physics, Hefei University of Technology, Hefei, China. [4]Shanghai Branch, Hefei National Laboratory, Shanghai, China. [5]Shanghai Research Center for Quantum Sciences, Shanghai, China. [6]Collaborative Innovation Center of Extreme Optics, Shanxi University, Taiyuan, China. [7]These authors contributed equally: Jinxian Guo, Chun-Hua Yuan, L. Q. Chen. ✉e-mail: jxguo@sjtu.edu.cn; chyuan@phy.ecnu.edu.cn; lqchen@phy.ecnu.edu.cn

## Results

### Experimental setup

To efficiently memorize hyper-dimensional information, it is necessary to be capable of coherently storing and retrieving OAM spatial modes and SAM polarization modes in the signal with an efficiency of above 90% and a fidelity exceeding the no-cloning limitation. Here, we experimentally demonstrate the memory ability of spatial and polarization information of current memory system. The experimental diagram is presented in Fig. 1a. An $^{87}$Rb atomic vapor cell with a length of 7.5 cm and a diameter of 2.5 cm is placed inside a cylindrical magnetic shielding to reduce the influence of surrounding magnetic fields. The atomic energy levels are shown in Fig. 1b. The atomic cell without buffer gas and anti-relaxation coating is heated to 84 °C to achieve the atomic density approximately $5.2 \times 10^{10}$ cm$^{-3}$. Before the write-in process, most atoms are populated on the hyperfine level $|m\rangle$ with approximately equal population distribution among Zeeman sublevels by the 30 $\mu$s-long pump pulse. The 20 ns-long signal beam, signed as $S_{in}$, is spatially overlapped with the strong write beam ($W$) by a polarization beam splitter (PBS), and then enters into the atomic cell. $S_{in}$ is partially stored as the spin excitation $S_a$ with write-in efficiency $\eta_W$ driven by the $W$ pulse. The rest signal leaks out of the atoms as $S_L$. After some delay time, the atomic excitation can be retrieved back as readout optical signal ($S_R$) by a strong read pulse ($R$) in the backward direction. $S_R$ and $S_{in}$ are spatially overlapped and split using Faraday rotator, half-wave plate and polarization beam splitter as shown in Fig. 1a. Finally, temporal waveforms and energies ($E_{in}$ and $E_R$) of the input $S_{in}$ and retrieved $S_R$ signals are measured at the end of the atomic cell by photo detector. The memory efficiency ($\eta_m$) is then calculated as $\eta_m = E_R/E_{in}$[19–21].

The vortex beam studied in this paper is a superposition of Laguerre-Gaussian (LG) modes with different OAMs and radial index $p = 0$[22], which can be written as:

$$LG_l(r, \phi, z) = E_l(r, \phi, z)\exp(il\phi), \tag{1}$$

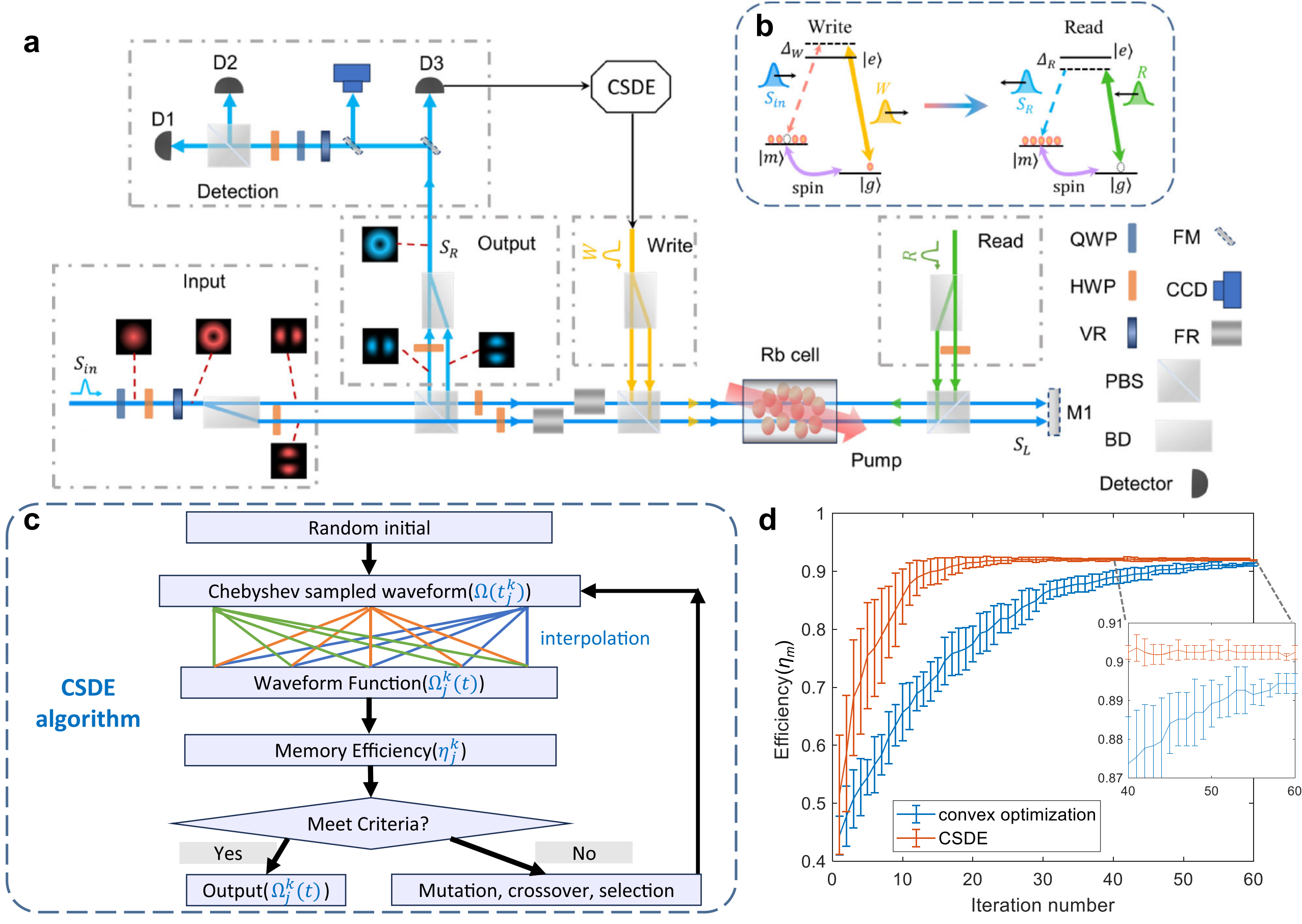

**Fig. 1 | Quantum memory scheme. a** Experimental setup. VR vortex retarder, HWP half-wave plate, QWP quarter-wave plate, BD beam displacer, FR Faraday rotator, FM Flip mirror, PBS polarization beam splitter, CCD charge coupled device camera, Rb cell $^{87}$Rb atomic vapor cell. $W$ write pulse, $R$ read pulse, $S_{in}$ input signal, $S_L$ leaked signal, $S_R$ retrieved signal, D1-D3 photo detector, M1 folding mirror. The powers of the $W/R$ beam is 270 mW and 180 mW, respectively. The pump and input signal are 30 $\mu$s and 20 ns long, respectively. **b** Atomic energy level and light frequencies. $|g\rangle$: $|5^2S_{1/2}, F = 1\rangle$ and $|m\rangle$: $|5^2S_{1/2}, F = 2\rangle$ are the two hyperfine ground states of $^{87}$Rb $D_1$-line; $|e\rangle$: $|5^2P_{1/2}, F = 2\rangle$ is the excited state; the detuning frequencies of $W$ and $R$ lights are $\Delta_W = 1.8$ GHz and $\Delta_R = -1.7$ GHz, respectively. $S_{in}$ and $W$ satisfy two-photon resonant condition. **c** The flowchart of AI algorithm, that is, Chebyshev sampling differential evolution (CSDE) algorithm, which is the mixing of Chebyshev sampling, residual network and convex optimization. The waveform of $W$ is optimized via CSDE algorithm. **d** Efficiency as a function of the iteration number of convex optimization and CSDE algorithm. Memory efficiency $\eta_m$ is the ratio of the average photon number of $S_R$ to that of $S_{in}$.

where $\phi$ is the azimuthal angle in the transverse plane, and $l$ is the topological charge. In our experiment, the vortex beam is generated by a light beam passing through a vortex retarder (VR). The SMG signal state $|\varphi(t)\rangle$, after carrying OAM information $|LG_l\rangle$ and SAM information $|P\rangle$, can be written as $|\Psi_l\rangle = |LG_{pl}\rangle \otimes |P\rangle \otimes |\varphi(t)\rangle$.

The SMG signal state $|\varphi(t)\rangle$ with a vertically linear polarization is injected into the VR, which can be decomposed into a coherent superposition of left- and right- circularly polarized (LCP/RCP) components with equal amplitude, acquires opposite signs of OAM through the action of VR. This generates an output superposition state $|\Psi_l\rangle$ with distinct OAM signatures:

$$|\Psi_l\rangle = \frac{1}{\sqrt{2}}\left(|LG_l\rangle \otimes |RCP\rangle + |LG_{-l}\rangle \otimes |LCP\rangle\right) \otimes |\varphi(t)\rangle. \tag{2}$$

This state is equivalent to:

$$|\Psi_l\rangle = \frac{1}{2}\left[\left(|LG_l\rangle + |LG_{-l}\rangle\right)|H\rangle + \left(|LG_l\rangle - |LG_{-l}\rangle\right)|V\rangle\right] \otimes |\varphi(t)\rangle. \tag{3}$$

The beam displacer (BD) splits the optical field into two spatially shifted paths based on polarization (H/V). Thus, the two mutually-displaced LG modes interfere coherently, generating two vector optical fields (VOFs) that exhibit petal-like intensity patterns, as shown in Fig. 2.

Prior to entering the atomic cell, both the $W$ and $R$ beams are split into two separate beams. The $W$ beams spatially overlap with two VOF signals, storing them as two spin excitations. Subsequently, two $R$ beams retrieve spin excitations into two VOF readout signals, which are then combined into one OAM beam upon exiting the cell via BD. Here, the topological charge $l$ of the OAM mode in signal $S_{in}$ varies from −5 to 5. The spatial pattern of OAM signal is measured using a CCD. The fidelity of OAM and SAM information can be analyzed via a detection system shown in of Fig. 1a. Both the $S_{in}$ and $S_R$ signals consist of three mutually independent components: $|\varphi(t)\rangle$, $|LG_l\rangle$, and $|P\rangle$. In efficiency optimization, we individually optimize the SMG, OAM and SAM parts to their respective maximum efficiencies.

### AI-optimized memory efficiency

We first maximize the memory efficiency for SMG signal by precisely matching the temporal waveform of the write pulse—a critical requirement for achieving high-efficiency quantum memory. Subsequently, we extend the optimized temporal waveform that maintain consistent memory performance across all topological charges and arbitrary polarization states. However, nonlinear waveform matching and the nonlinear response of modulators make the optimization complex and challenging. Although convex optimization[23] and Gaussian approximations can provide quick approximate solutions for certain cases, such as input signals with Gaussian temporal shape, they fall short when it comes to arbitrary waveforms. For practical quantum networks, the optimization method should be adapted to any input waveform, ensuring optimal performance across a wide range of conditions. The existing algorithms face slow convergence and the problem of getting stuck in local minima when searching for the optimal waveform in a vast search space, limiting their practicality. To address these challenges, we develop a waveform optimization method based on a modified DE algorithm using Chebyshev sampling. This approach integrates sparse Chebyshev node parameterization (typically 20 nodes) with a neural network-assisted waveform generator to address both global convergence and local nonlinear compensation. The DE algorithm encodes solutions as Chebyshev sampling points that strategically capture critical waveform features, significantly reducing the parameter space dimensionality compared to dense-sampling methods. A pre-trained residual network (20-50-30 architecture with ReLU-activated hidden layers) is cascaded after spline interpolation to reconstructs missing high-frequency components through learned prior knowledge. The network, trained on 10,000 waveforms via Adam optimization (learning rate $10^{-4}$, 100 epochs), learns to predict nonlinear corrections through end-to-end waveform mapping. During optimization, DE performs global exploration by evolving Chebyshev node parameters through differential mutation and crossover. By employing this method shown in Fig. 1c, we can significantly reduce the complexity of the parameter space. This enables high-precision waveform optimization for arbitrary input signals in a relatively short time. Experimentally, $|\varphi(t)\rangle$ is a coherent signal with 20 ns-long Gaussian temporal shape and SMG spatial mode. As shown in Fig. 1d, the memory efficiency of the $S_{in}$ signal is stably achieved to 92%, which support efficient quantum memory of hyper-dimensional signal well.

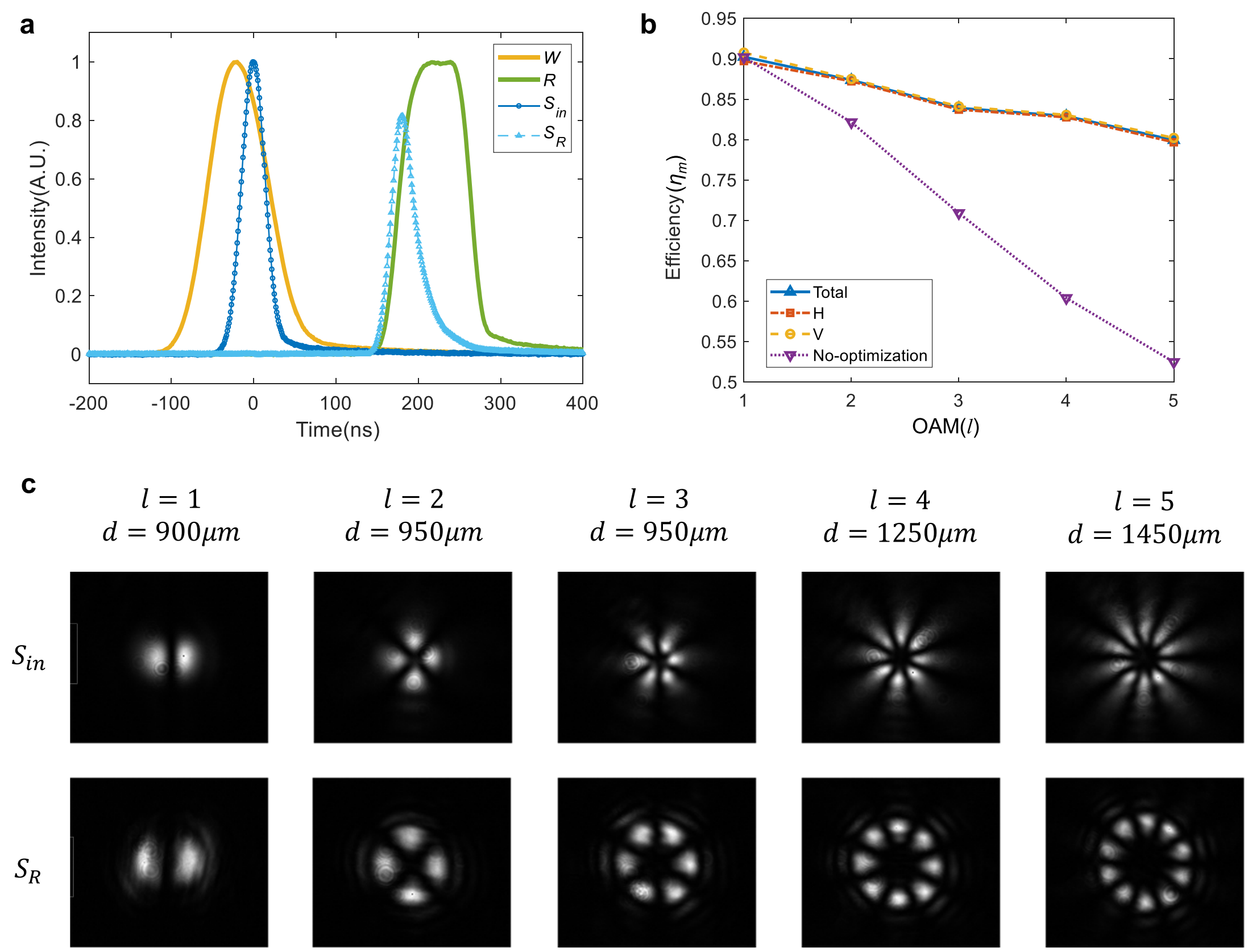

**Fig. 2 | Experimental efficiency of hyper-dimensional broadband memory.** **a** Temporal waveforms of the input ($S_{in}$, deep blue), write ($W$, yellow), the retrieved ($S_R$, sky blue) and read ($R$, green) pulses in one-time memory. The $S_{in}$ signal pulse is 20 ns, corresponding to a bandwidth of 50 MHz. **b** Memory efficiency $\eta_m$ as a function of topological charge $l$ of OAM, the blue triangle is the overall efficiency, the orange square and yellow circle represent the horizontal and vertical polarization path, these three curves exhibit overlap with each other. The purple inverted triangle represents the efficiency without spatial optimization. **c** The spatial patterns of $S_{in}$ at the center of the atomic cell and $S_R$ with topological charge $l$ from 1 to 5.

Then, we optimize the memory efficiency of hyper-dimensional signals based on the temporal waveforms of the input $S_{in}$ pulse, optimized $W$ pulse, the $R$ pulse and retrieved $S_R$ signal shown in Fig. 2a. The efficiency of different OAM modes is shown in Fig. 2b. In the case of topological charge $l$ = 1, −1, efficiency can reach 91%, while gradually dropping to 80% at $l$ = 5, −5. The decrease of $\eta_m$ is resulted from the increase of the beam waist as $l$. The beam waist $\omega$ of the OAM mode is $\omega = \sqrt{l+1}\omega_0$, where $\omega_0$ is the single-mode Gaussian beam waist[14]. The beam waist and central singular point both increase as $l$, which is clearly shown in Fig. 2c. Larger waist of $S_{in}$ requires larger diameter of the $W$ beam leading to a decrease in the light density, resulting in a reduction in atom-light coupling strength. This is the main reason for efficiency decrease of higher-order OAM signals. Therefore, reducing the beam waist of the OAM signal and the $W$ beam using lens is the effective way to increase the coupling coefficient and improve memory efficiency. However, smaller diameter and larger topological charge also simultaneously reduces the number of the effective atoms, which in turn leads to a weakening of the coupling strength. Ultimately, the memory efficiency is the result of optimal coupling determined by both the number of atoms and the waist of optical fields. As shown in Fig. 2b, after optimizing the spot size of OAM signal, the decline rate of memory efficiency is greatly alleviated.

## Fidelity

Fidelity is the core criterion for judging whether it is quantum memory. For both input and readout signals comprising three independent components, $|\Psi_l\rangle = |LG_l\rangle \otimes |P\rangle \otimes |\varphi(t)\rangle$, the total memory fidelity follows the relation $F = F_{\mathrm{SMG}} \times F_{\mathrm{OAM}} \times F_{\mathrm{SAM}}$, where $F_{\mathrm{SMG}}$, $F_{\mathrm{OAM}}$, and $F_{\mathrm{SAM}}$ represent the fidelity of the SMG signal, OAM and SAM information, respectively. We independently measured these fidelity parameters, with results presented in Fig. 3. First, we measured the $F_{\mathrm{SMG}}$, which is calculated based on the density matrix $\rho_{in}$ and $\rho_R$ of $S_{in}$ and $S_R$ measured using homodyne detection (details see methods "The fidelity of SMG")[24] via

$$F(\rho_{in}, \rho_R) = \mathrm{Tr}\left[\sqrt{\sqrt{\rho_{in}}\rho_R\sqrt{\rho_{in}}}\right]^2. \tag{4}$$

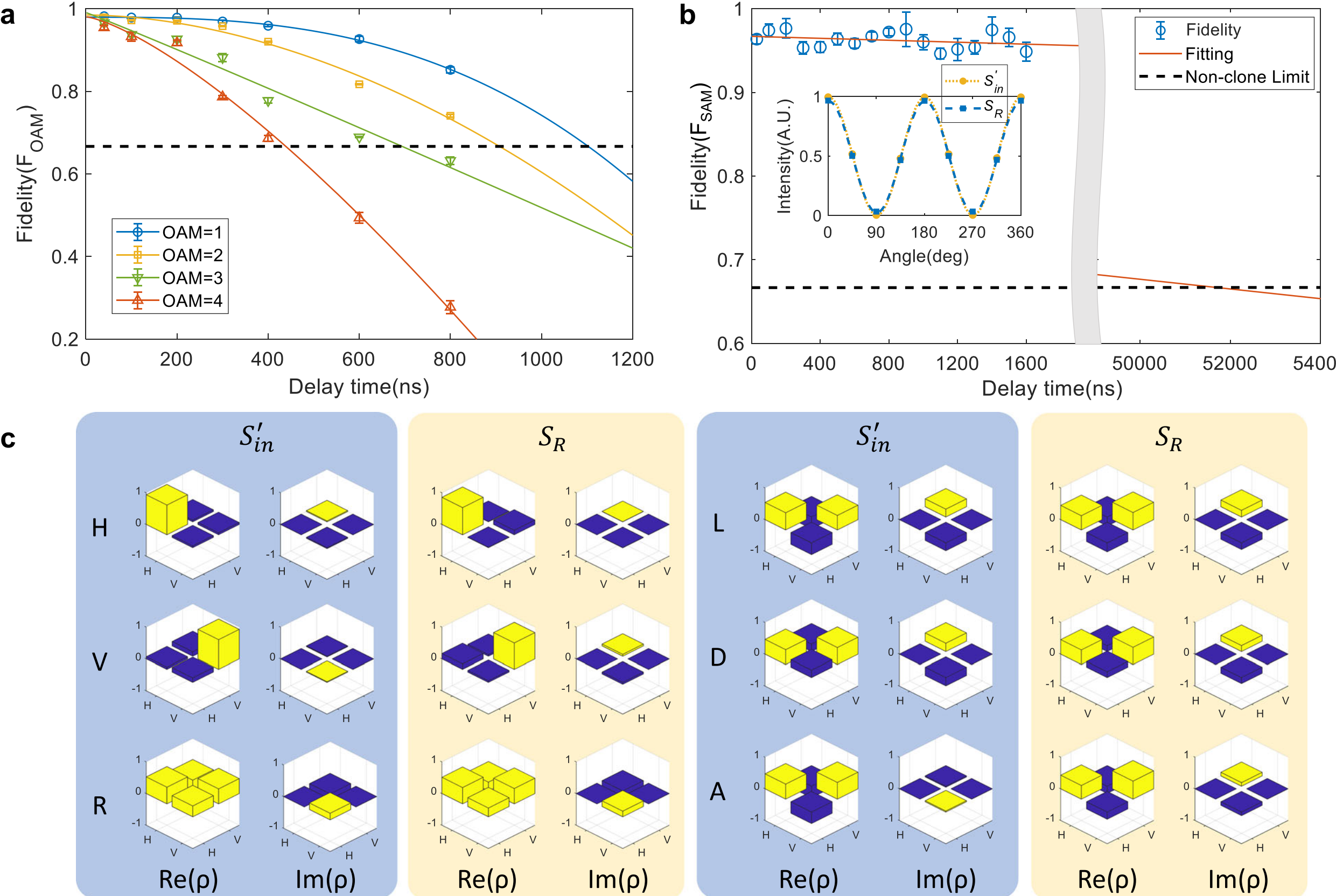

**Fig. 3 | Fidelity data of AI-optimized memory.** Fidelity as a function of the delay time for OAM (**a**) and SAM (**b**). The delay time is the time difference between $W$ and $R$ pulses, the error bars are the standard deviation of 10 consecutive acquisitions. The inset of (**b**): intensity of horizontal polarization of before and after memory as the polarization angle changes from 0 to $2\pi$. **c** The projection results of SAM information encode on OAM ($l$ = 2) of $S'_{in}$ and $S_R$ signals. $S'_{in}$ is the $S_{in}$ signal passing through the whole optical path without memory by blocking pump field.

The four-wave mixing noise in current memory process is very low, owing to both the far-off resonance condition of Raman memory and the reverse readout scheme adopted in this experiment. Thus, the memory fidelity of SMG ($F_{\mathrm{SMG}}$) is achieved up to 99% at an average photon number of 0.9 photon/pulse.

Then we measure the $F_{\mathrm{OAM}}$. Figure 2c shows the spatial patterns of the OAM signals, $S_{in}$ and $S_R$. By counting the number of light spots in the spatial patterns of the $S_{in}$ and $S_R$ signals, we can achieve integer-order OAM information. From the patterns in Fig. 2c, the topological number of the OAM pattern $l$ is 1-5 from left to right. By analyzing the azimuth angle of the input and output optical field spatial patterns through image recognition, we can obtain the shift in azimuth angle and the fidelity of the OAM memory. After a delay time of 40 ns, the azimuth angle shift of the OAM spatial pattern is smaller than $10^{-2}$ rad for $l = 1$ and $10^{-3}$ rad for $l = 4$, demonstrating that current atomic memory can store the spatial patterns information of OAM signal well. After the BD, two VOA signals show petal-like patterns, which include the topological charge $l$ and initial azimuthal phase $\varphi_0$ information[25]. We extract the information by projecting azimuthal interference pattern onto the basis of $e^{-il\varphi}$ to achieve the density matrix $\rho(l, \varphi_0)$ of the OAM signal. According to the quantum fidelity of Eq. (4), the fidelity $F_{\mathrm{OAM}}$ can be calculated. Figure 3a shows the fidelity of OAM as a function of delay time between the $W$ and $R$ pulses. The fidelity $F_{\mathrm{OAM}}$ is 95.5%~98.1% when delay time $\tau = 40$ ns, and decreases with the topological charge and delay time. We define the memory time for OAM information is the delay time when the fidelity decrease to the no-cloning limit 67%[26]. The memory time is shortest 0.42 $\mu$s for $l = 4$ and longest 1.1 $\mu$s for $l = 1$, corresponding time-bandwidth product of 21 and 55. In current memory, the diffusion due to atomic thermal motion is the main impact factor on the memory time of OAM spatial patterns. Our experimental results show that higher-order OAM modes experience stronger diffusion effects, as both the beam waist and central singularity scale with the topological charge $l$, consistent with diffusion theory[27].

According to Fig. 2b, the memory efficiency of the OAM information decreases with the topological charge even after optimizing the spot size. Therefore, efficiency for high-order OAM signal may be very low. This is also why SAM information is encoded with the OAM, which can ensure both high dimensionality and high memory efficiency. In experiment, when the OAM mode is encoded on the $S_{in}$ signal by VR, the SAM information can be simultaneously encoded on $S_{in}$ signal using wave plates (HWP, QWP) as shown in Fig. 1a. After the signals are read out as $S_R$, they are converted back to polarization states by a second VR plate in the detection optical path. The SAM encoded on signal introduces two more dimensions independent of the OAM, namely the horizontal and vertical polarization (H, V). By combining H and V polarization, SAM information can be added on the OAM. Such operation expands the dimensional state number from 11 ($l = -5 \sim 5$) to 22.

We experimentally demonstrate the quantum memory of SAM states encoded on OAM with $l = 2$. The fidelity of SAM information is measured and presented in Fig. 3b. Before measuring the memory fidelity, we need to assess the impact of optical components in the signal propagation path on SAM information. We qualify it by the visibility of $S_{in}$ and $S'_{in}$ as the inset of Fig. 3b shown, which is respective 99% and 93%, demonstrating that the optical components in the optical path cause a certain degree of damage to the fidelity of SAM information. $S'_{in}$ is the $S_{in}$ signal passing through the whole optical path without memory by blocking pump field. Then the memory fidelity of the SAM information is measured and analyzed by comparing the $S_{in}$ and $S_R$ signals. By using QWP and HWP, we can obtain six polarization states, that is, $|H\rangle = |0\rangle$ and $|V\rangle = |1\rangle$, $|D\rangle = |0\rangle + |1\rangle$ and $|A\rangle = |0\rangle - |1\rangle$ by adding a polarization angle of 45° via HWP, $|R\rangle = |0\rangle + i|1\rangle$ and $|L\rangle = |0\rangle - i|1\rangle$ by adding a phase difference of $\pi/2$ via QWP. The states $|D\rangle$, $|A\rangle$, $|R\rangle$ and $|L\rangle$ are the superposition states of $|H\rangle$ and $|V\rangle$.

We measured six polarization projections of $S'_{in}$ and $S_R$ signals to experimentally reconstruct the density matrix of the polarization states by quantum state tomography (QST) performed using a combination of a QWP, a HWP, a PBS and photo detector modules (D1 and D2) as shown in Fig. 1a. This fidelity reflects the preservation of SAM information during the memory process, without considering the influence of optical components on SAM information. The density matrix results for the six polarization states are shown in Fig. 3c. Based on the density matrices $\rho_{S_{in}}$ and $\rho_{S_R}$, we obtained the fidelity values $F_{\mathrm{SAM}}$. The memory fidelity $F_{\mathrm{SAM}}$ for six polarization states $|H\rangle$, $|V\rangle$, $|R\rangle$, $|L\rangle$, $|D\rangle$ and $|A\rangle$) are 98.9%, 97.4%, 99.5%, 98.3%, 99.3%, and 97.6%. The whole fidelity $F = F_{\mathrm{SMG}} \times F_{\mathrm{OAM}} \times F_{\mathrm{SAM}}$ ranges from a minimum value of 92% to a maximum value of 95%, which are far beyond no-cloning limitation 67%[26], indicating that current hyper-dimensional memory is quantum one[28]. Figure 3b presents average polarization fidelity $F_{\mathrm{SAM}}$ as a function of the delay time. The memory time is approximately 50 $\mu$s, which is much longer than that of OAM one. We achieve high-fidelity SAM memory. For signal carrying both OAM and SAM, the final memory time depends on the shorter one, yielding a 0.4 $\mu$s memory time for current hyper-dimensional memory. In future, the memory time can be effectively provided by introducing buffer gas into the system and employing reading light with an expanded beam profile.

## Discussion

We have demonstrated a hyper-dimensional and high bandwidth quantum memory in an $^{87}$Rb atomic vapor cell with an efficiency of 91%, a bandwidth of 50 MHz, and the fidelity of up to 95.5% for the OAM information, 97.4% for the SAM information and 92% for whole hyper-dimensional signal. This is the highest efficiency to date for both high bandwidth and dimensional quantum memory. In this paper, we successfully realize high-efficiency, high-fidelity broadband quantum memory of the states which are encoded by the 11 OAM and 2 polarization basis vectors, demonstrating that current memory can potentially store up to 22-dimensional states if non-separable superpositions are generated and preserved. Our work undoubtedly provides strong support for the realization of large-scale and high-speed quantum networks.

## Methods

### Frequency control of the optical field

The memory process imposes stringent frequency requirements on all optical fields. First, the frequency of the input signal and write fields must satisfy the two-photon resonance condition. And furthermore, as the write beam is a strong control field, which induces an AC Stark shift ($\Delta_{AC}$) on atom-light interaction. Consequently, the frequency difference between input signal and write beams must be stabilized at $\Delta_{hf} + \Delta_{AC}$, where $\Delta_{hf} =$ 6.834 GHz corresponds to the $|g\rangle \rightarrow |m\rangle$ hyperfine transition in the $^{87}$Rb atoms. The AC Stark shift $\Delta_{AC}$ depends on the intensity of write field, we operate a real-time adjustment through a feedback loop to compensate for intensity fluctuations. In order to achieve stable frequency output, an optical phase-locked loop is used to lock the frequencies of the $S_{in}$ and write lasers in the experiment. Second, there is a frequency difference between input signal and retrieved signal, that is, $\Delta_W + \Delta_R$. $\Delta_W$ and $\Delta_R$ are the single-photon detuning of the write and read light, respectively. $\Delta_W = 1.8$ GHz and $\Delta_R =$ $-1.7$ GHz under optimal conditions, so the frequency difference between $S_R$ and $S_{in}$ is 3.5 GHz. To enable homodyne detection of the retrieved signal ($S_R$), we generate a local oscillator field by downshifting the $S_{in}$ laser by exactly 3.5 GHz using an electro-optic modulator (EOM).

### The flowchart of AI algorithm

To find the best waveform parameter $\Omega(t)$ for $W$ pulse. The Chebyshev sampling differential evolution algorithm initiates by determining the degree of Chebyshev polynomial to calculate the corresponding time nodes $t_j$. The first population $\Omega(t_j^1)$ is randomly generated, followed by spline interpolation at the Chebyshev nodes, followed by spline interpolation to create smooth waveform functions $\Omega_j^1(t)$ corresponding to the optical field temporal waveform. Then, applying a nonlinear correction on $\Omega_j^1(t)$ by a pre-trained residual network, to reconstruct missing high-frequency components through learned prior knowledge. After loading the modulation frequency of the acoustic optic modulator (AOM) on the waveform functions $\Omega_j^1(t)$, use an arbitrary waveform generator to generate the corresponding electrical signal, and driving the AOM to experimentally output the waveform of $W$ pulse. The algorithm calculates memory efficiency $\eta_j^1$ for each waveform $\Omega_j^1(t)$ and applies differential evolution operations, including mutation, crossover, and selection[29], to refine the population $\Omega(t_j^1)$. This process continues until a convergence criterion is met, at which point the waveform parameters $\Omega(t_j^k)$ that maximize efficiency are identified as the optimal solution.

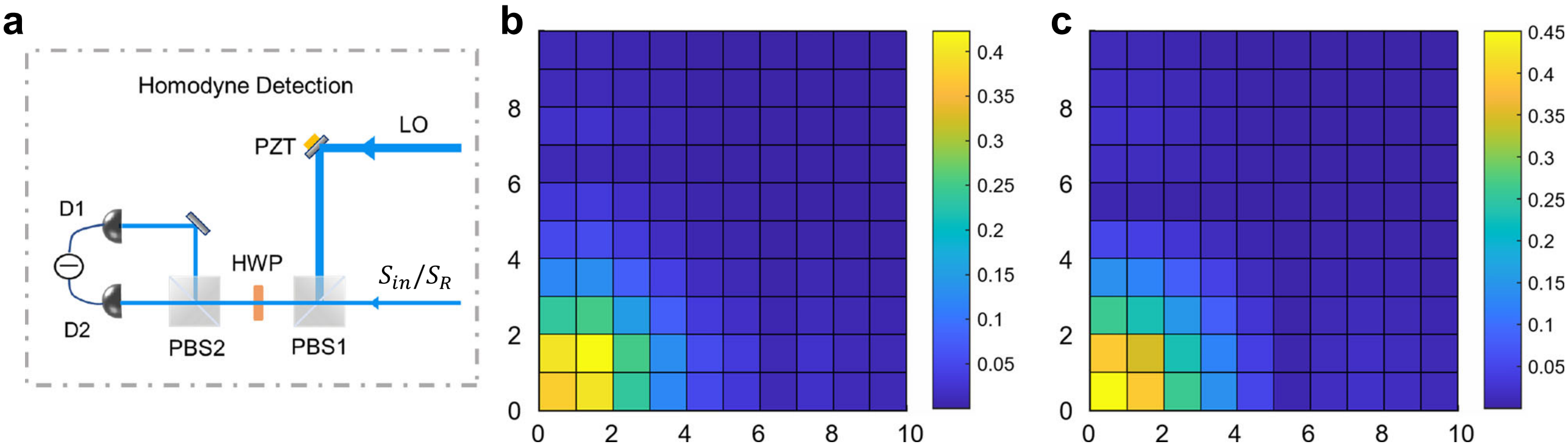

**Fig. 4 | The density matrices of the SMG memory. a** Experimental setup for homodyne measurement. PZT piezoelectric transducer, LO local oscillator, D1 and D2 photodiode, The density matrices of the input (**b**) and output (**c**) signal pulses with 0.9 photon/pulse on average.

### The fidelity of SMG

The fidelity of SMG is achieved using the equation $F(\rho_{in}, \rho_R) = \mathrm{Tr}\left[\sqrt{\sqrt{\rho_{in}}\rho_R\sqrt{\rho_{in}}}\right]^2$, where $\rho_{in}$ and $\rho_R$ are the reconstructed density matrices of $S_{in}$ and $S_R$ with SMG mode, respectively. The setup for homodyne detection is shown in Fig. 4a. Local oscillators with the same frequency as the signal pulses $S_{in}$ and $S_R$ for homodyne measurement are prepared by the method described in subsection "Frequency control of the optical field". In experiment, we record $5 \times 10^3$ sets of quadrature amplitudes of the $S_{in}$ and $S_R$ pulses while varying the phase of the local oscillator between 0 and $2\pi$ by scanning the piezoelectric transducer, and then reconstruct the density matrices by tomographic reconstruction[30]. The density matrix elements of the $S_{in}$ and $S_R$ pulses are obtained based on the quadrature-amplitude results using the maximum-likelihood reconstruction method[31]. The results are plotted in Fig. 4b, c, with the input pulses at an average photon number of 0.9 photon/pulse, the fidelity of SMG memory $F_{\mathrm{SMG}} = 99\%$.

## Data availability

All relevant data and figures supporting the main conclusions of the document are available on request. Please refer to Zeliang Wu at zlwu@phy.ecnu.edu.cn.

## Code availability

All relevant code supporting the document is available upon request. Please refer to Zeliang Wu at zlwu@phy.ecnu.edu.cn.

## Acknowledgements

This work is supported by the National Natural Science Foundation of China Grants No. U23A2075, No. 12274132, No. 12374331, No. 11904227, No. 12104161, No. 12304391 and No. 11974111; Fundamental Research Funds for the Central Universities; theChina Postdoctoral Science Foundation (Grant No. 2023M741187, No. GZC20230815); the Innovation Program for Quantum Science and Technology (Grant No. 2021ZD0303201); the Shanghai Municipal Science and Technology Major Project (Grant NO. 2019SHZDZX01); the Innovation Program of Shanghai Municipal Education Commission (Grant No. 202101070008E00099), and the Shanghai Science and Technology Innovation Project (Grant No. 24LZ1400600). W.Z. also acknowledges additional support from the Shanghai talent program.

## Author contributions

Z.L. and L.Q.C. performed the measurements and analysis of the results. Z.L. and J.G. designed the device. Z.L., Z.Y. and H.W. contributed to the implementation of the experimental system. L.Q.C., C.Y., Z.L., Z.Y. and J.G. developed the idea and carried out theoretical simulations of the system. Z.L. and J.G. developed the artificial intelligent algorithm. Z.L., L.Q.C., J.G. and Z.Y. wrote the manuscript. Z.L., L.Q.C., C.Y., H.W. and W.Z. edited the manuscript. L.Q.C. provided supervision and guidance during the project. All authors contributed to the discussions and interpretations of the results.

## Competing interests

The authors declare no competing interests.

## Additional information

**Correspondence** and requests for materials should be addressed to Jinxian Guo, Chun-Hua Yuan or L. Q. Chen.